\documentclass[3p,times]{elsarticle}

\usepackage{ecrc}
\usepackage{subfigure} 		
\usepackage{graphicx}

\volume{00}

\firstpage{1}

\journalname{Nuclear Physics A}

\runauth{L. Apolin\'{a}rio, N. Armesto and L. Cunqueiro}

\jid{nupha}

\usepackage{amssymb}

\usepackage[figuresright]{rotating}

\begin{document}

\begin{frontmatter}




\title{Background subtraction and jet quenching on jet reconstruction}


\author[label1,label2]{Liliana Apolin\'{a}rio}
\ead{lilianamarisa.cunha@usc.es}
\author[label1]{N\'{e}stor Armesto}
\author[label3,label4]{Let\'{i}cia Cunqueiro}

\address[label1]{Departamento de F\'{\i}sica de Part\'{\i}culas and  IGFAE, Universidade de Santiago de Compostela, 15782 Santiago de Compostela, Galicia-Spain}
\address[label2]{CENTRA, Instituto Superior T\'ecnico, Universidade T\'ecnica de Lisboa, Av. Rovisco Pais, 1049-001, Lisboa, Portugal}
\address[label3]{Laboratori Nazionali di Frascati, INFN, Via E. Fermi, 40, 00044 Frascati (Roma), Italy}
\address[label4]{PH Department, CERN, CH-1211 Gen\`eve 23, Switzerland}

\begin{abstract}
In order to assess the ability of jet observables to constrain the characteristics of the medium produced in heavy-ion collisions at the LHC, we investigate the influence of background subtraction and jet quenching on jet reconstruction, with focus on the dijet asymmetry as currently studied by ATLAS and CMS. Using a toy model, we examine the influence of different background subtraction methods on dijet momentum imbalance and azimuthal distributions. We compare the usual jet-area based background subtraction technique and a variant of the noise-pedestal subtraction method used by CMS. The purpose of this work is to understand what are the differences between the two techniques, given the same event configuration. We analyze the influence of the quenching effect using the Q-PYTHIA Monte Carlo on the previous observables and to what extent Q-PYTHIA is able to reproduce the CMS data for the average missing transverse momentum that seems to indicate the presence of large angle emission of soft particles.
\end{abstract}


\end{frontmatter}


\section{Introduction}
\label{Intro}

\par One of the main subjects in heavy-ion collisions at ultra-relativistic high energy is the characterization of the produced medium through hard probes, like the suppression of high momentum particles or jet-like correlations. With the advent of the LHC, jet observables\cite{Aad:2010bu,Chatrchyan:2011sx,Chatrchyan:2012nia} have also moved to the focus on the discussion. In order to extract precisely the medium properties, the analysis of jet production requires several components: generation of medium-modified jet events together with a realistic background; and reconstruction of jets and background subtraction as close as possible to the experimental analysis. But this program faces several difficulties, like how to model and generate a realistic fluctuating background and how to treat the coupling between the hard event and the background. Additionally, one needs to understand the different ingredients used in a jet analysis and how they affect the different observables. In this work, we want to investigate the effect of background subtraction and of quenching on several jet observables. For that, we use Q-PYTHIA\cite{Armesto:2009fj} jets embedded in a simulated background, which follows a thermal plus power law $p_\perp$ spectrum. In order to easily tune the background parameters, we choose to use a toy model rather than a more realistic one. This paper is structured as follows: in section \ref{Procedure}, a summary of the background model, jet reconstruction and jet subtraction methods is provided. In section \ref{Results} a comparison between the different jet subtraction techniques for different jet observables is shown, followed by the effect of a jet quenching model, as implemented in the Q-PYTHIA Monte Carlo in section \ref{Quenching}. Finally, the conclusions are presented in section \ref{Conclusions}.

\section{Background Model and Jet Reconstruction}
\label{Procedure}

\par The background model that is used in this study is based on a thermal spectrum, where the inverse slope, $T$, the only free parameter of this model controls the exponential decay of soft particles. The hard part of the spectrum is constrained by a power law, $p_T^{-6}$. Particles are generated uniformly in $(\eta, \phi)$ and the multiplicity per pseudorapidity unit is fixed to $2100 \pm 5 \%$ particles. The temperature is chosen so that the region-to-region background fluctuations are of similar level as the ones measures by ALICE\cite{Abelev:2012ej} (corrected to include neutral particles), although the average level of contamination is somehow larger. Hence, the chosen values were $T = 0.7$, $T = 0.9$ and $T = 1.2$ GeV, that correspond to an average level of fluctuations of $7.69$, $10.75$ and $15.14$, respectively.  For jet reconstruction we use the anti-kt algorithm with a radius $R = 0.3$. For the jet background subtraction we compare the standard FastJet\cite{Cacciari:2011ma} method based on jet areas using the kt-algorithm with an $R = 0.4$ over a full stripe in $|\eta| < 2$, and a variant of the CMS technique\cite{Kodolova:2007hd}, which is based on a "noise/pedestal" subtraction. In this case, the phase space is divided according to the CMS calorimeter segmentation. Then, an average transverse energy, $< E_T^{tower} (\eta) >$ and a dispersion $\sigma_T^{tower} (\eta)$ is computed for each stripe in $\eta$. To each cell, the corresponding $< E_T^{tower} (\eta) >$ and $\sigma_T^{tower} (\eta)$ is subtracted and set to zero in case of a negative value. Using only particles that are in positive cells, jets are reconstructed with a minimum transverse energy above a cut, $E_{Tjets}$. All towers that are contained inside this list of jets are excluded and new background parameters are computed with the remaining list of particles. These new parameters are the final background estimation values used to perform a final correction to the initial transverse energy of each cell. Again, using only non-zero towers, the final list of jets is found. Since the final estimation of parameters should be the close as possible of the true background parameters, the parameter $E_{Tjets}$, that according to CMS is $10$ GeV, changes with the temperature, in the case of our toy model. In the end, the best estimation of the background parameters is given by $E_{Tjets} = 40$ for $T = 0.7$, $E_{Tjets} = 60$ for $T = 0.9$ and $E_{Tjets} = 70$ GeV for $T = 1.2$ GeV.

\section{Jet Subtraction on Jet Observables}
\label{Results}

\begin{figure}[h]
	\begin{center}
		\subfigure[Dijet Asymmetry using jet area method for background subtraction.]{
		\includegraphics[width=0.3\textwidth]{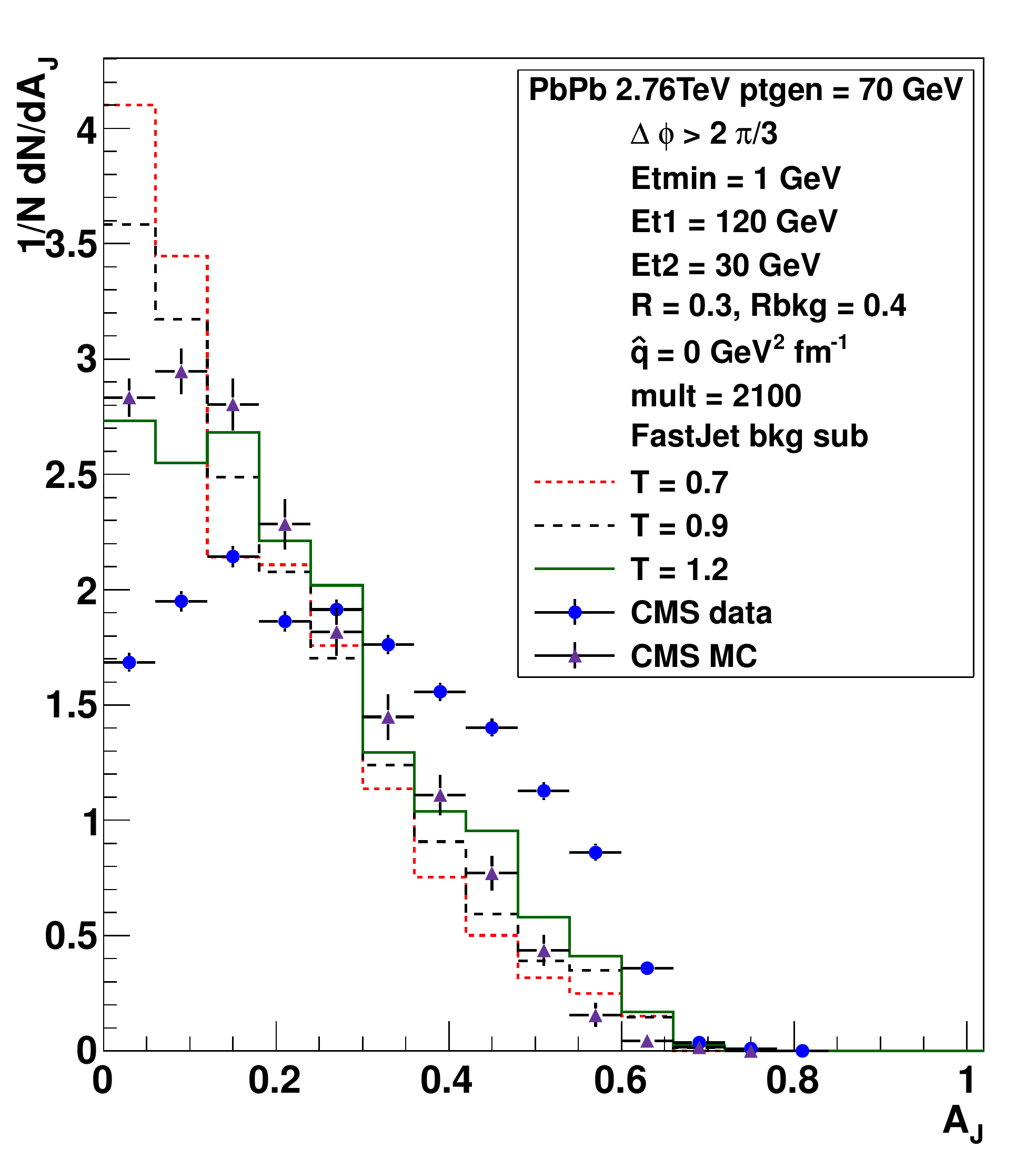}
		\label{fig:Aj_FJ_Fluct}
		}
		\subfigure[Dijet Azimuthal Correlation using jet area method for background subtraction.]{
		\includegraphics[width=0.3\textwidth]{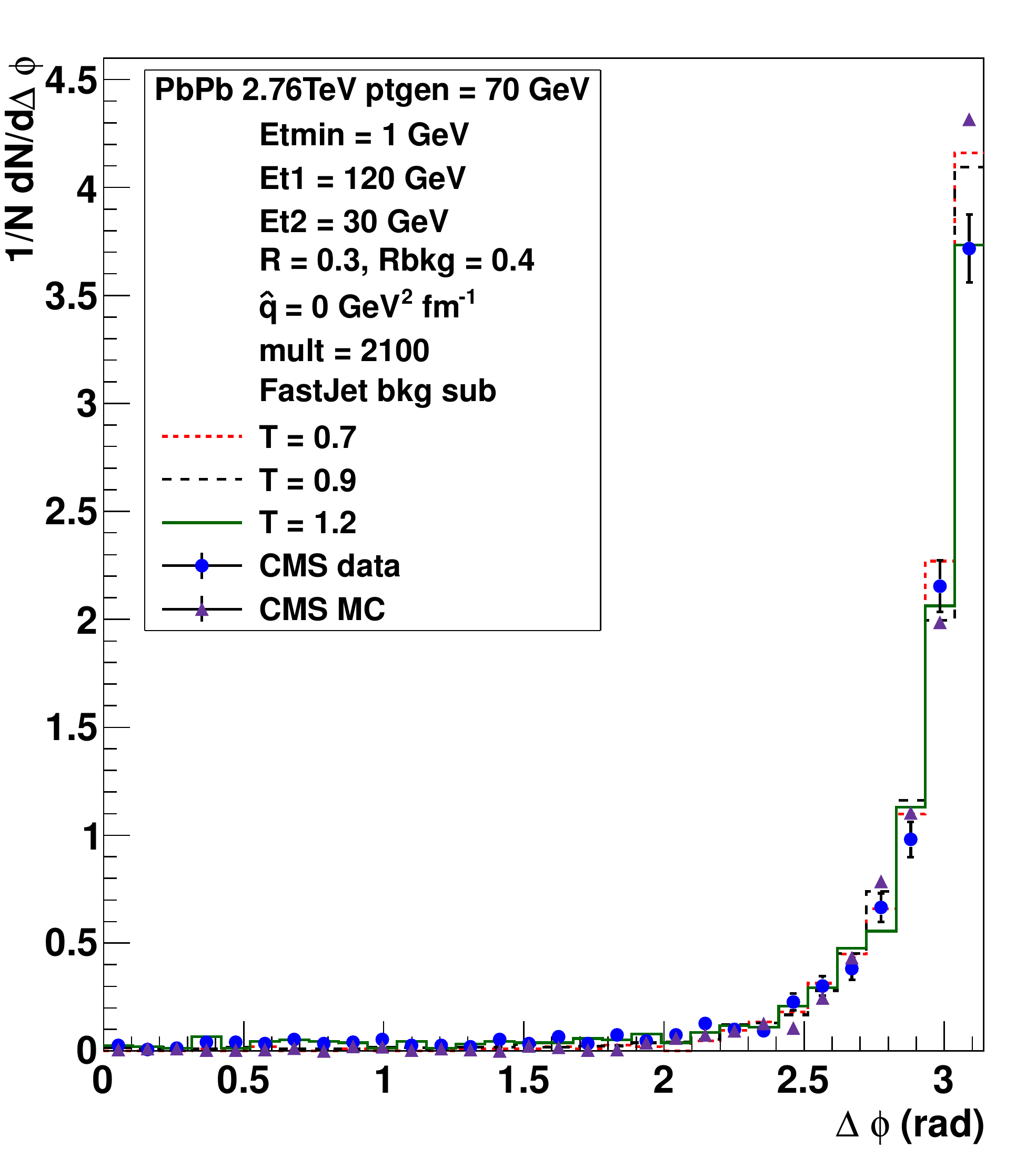}
		\label{fig:Dphi_FJ_Fluct}
		}
		\subfigure[Dijet Azimuthal Correlation using CMS-\emph{like} method for background subtraction.]{
		\includegraphics[width=0.3\textwidth]{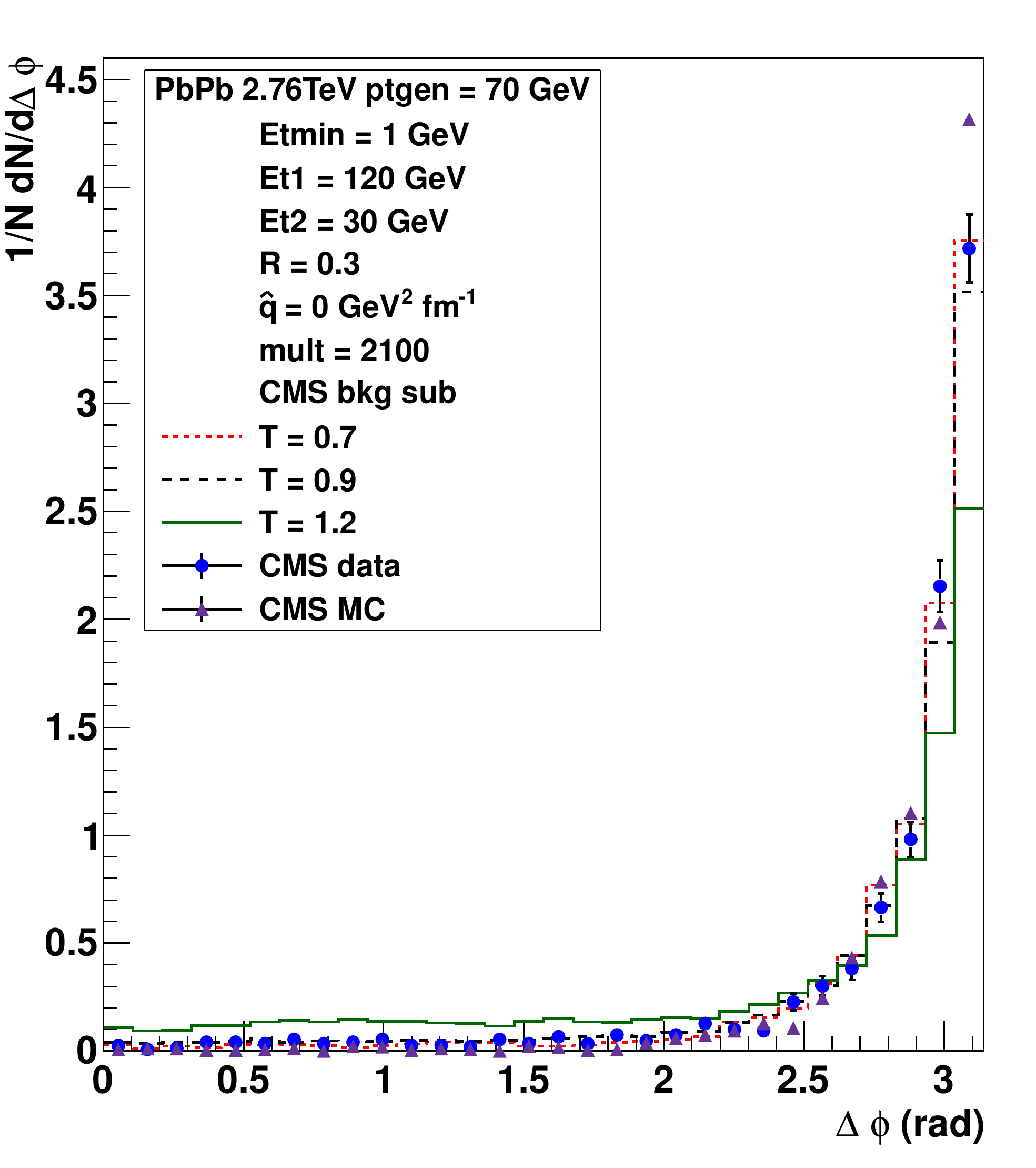}
		\label{fig:Dphi_CMS_Fluct}
		}
		\vskip-0.3cm
		\caption{Jet Observables for a simulation using Q-PYTHIA with a $\hat{q} = 0$ GeV$^2$ fm$^{-1}$ embedded in a background with different temperature parameters. The red/dotted curve corresponds to a $T = 0.7$, the black/dashed to a $T = 0.9$ and the green/solid to a $T = 1.2$. The blue points are the CMS data and the purple triangles the CMS Monte Carlo simulation.}
		\label{fig:Fluct}
	\end{center}
\end{figure}
\par In this section, our aim is to compare both background subtraction techniques that were outlined in the previous section, and to understand what is the effect of the background parameters, specifically the fluctuations, on the observables that we want to describe, dijet asymmetry and dijet azimuthal correlation. Here we present a comparison between the two background subtraction methods: CMS-\emph{like} and FastJet (jet areas). In figure \ref{fig:Aj_FJ_Fluct}, it is shown the effect of the fluctuations on the dijet asymmetry when using a background subtraction based on jet areas and as one can observe the effect of the fluctuations goes in the same direction than data, but no "realistic" values can account for the large asymmetry indicated by CMS data. On the contrary, the angular reconstruction of the jets seems to be unaltered, as one can see from figure \ref{fig:Dphi_FJ_Fluct}. As for the CMS-\emph{like} method, by construction, the resulting dijet asymmetry has a much smaller sensitivity to fluctuations, since the jet energy reconstruction is more accurate, but presents a larger deviation for the azimuthal correlation (figure \ref{fig:Dphi_CMS_Fluct}). 

\section{Quenching on Jet Observables}
\label{Quenching}

\par
\begin{figure}[h]
	\begin{center}
		\subfigure[Dijet Asymmetry.]{
		\includegraphics[width=0.3\textwidth]{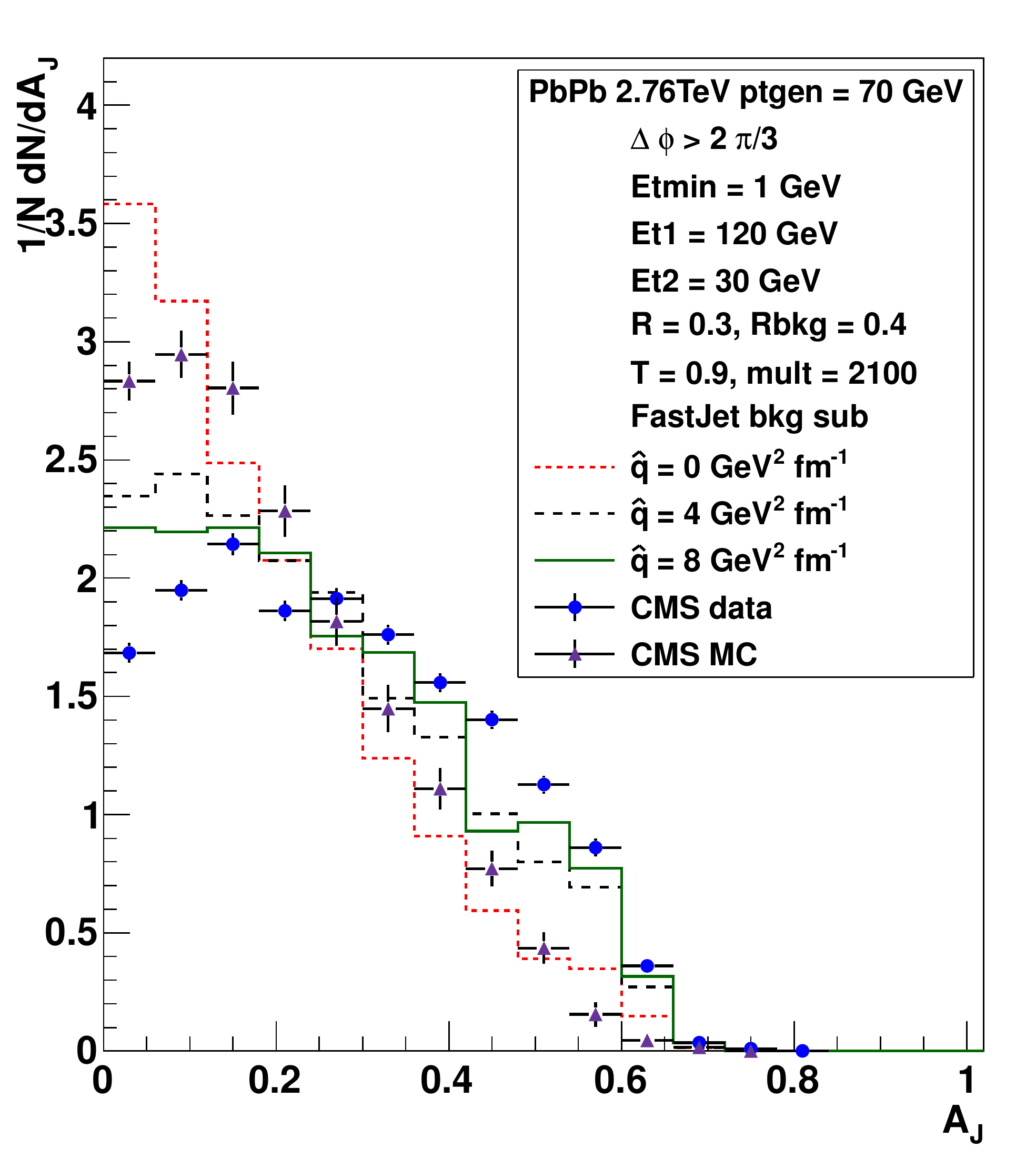}
		\label{fig:Aj_FJ_Quench}
		}
		\subfigure[Dijet Azimuthal Correlation.]{
		\includegraphics[width=0.3\textwidth]{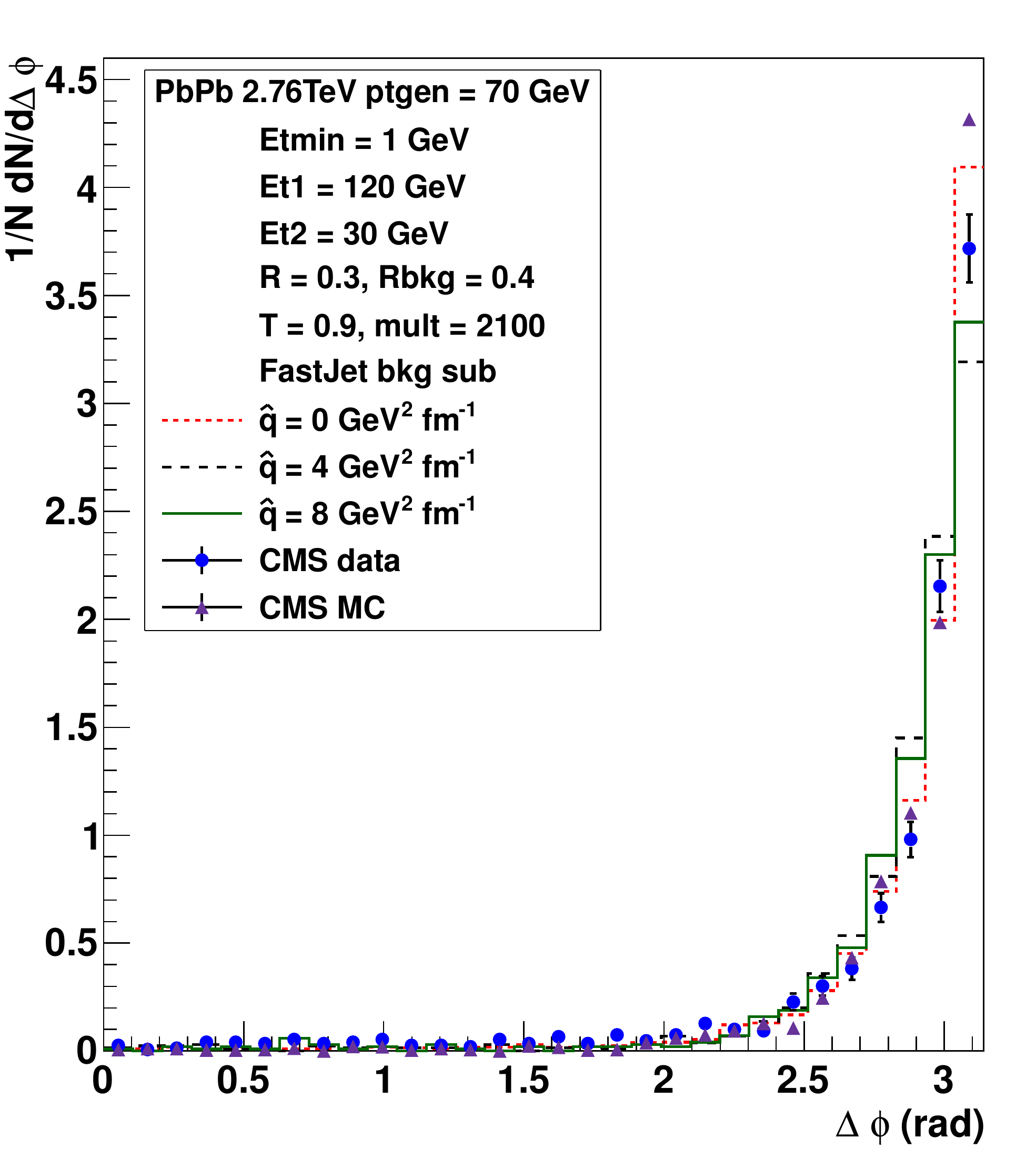}
		\label{fig:Dphi_FJ_Quench}
		}
		\vskip-0.3cm
		\caption{Jet Observables for a simulation using Q-PYTHIA with different $\hat{q}$ parameters, embedded in a background with $T = 0.9$ ($\sigma \sim 11$ GeV). The red/dotted curve corresponds to a $\hat{q} = 0$ GeV$^2$ fm$^{-1}$, the black/dashed to a $\hat{q} = 4$ GeV$^2$ fm$^{-1}$ and the green/solid to a $\hat{q} = 8$ GeV$^2$ fm$^{-1}$. The blue points are the CMS data while the purple triangles the CMS Monte Carlo simulation. The jet area method is used to subtract the background.}
		\label{fig:FJ_Fluct}
	\end{center}
\end{figure}

\begin{figure}[h]
	\begin{center}
		\subfigure[$< p_T^\parallel >$  with $\hat{q} = 0$ GeV$^2$ fm$^{-1}$.]{
		\includegraphics[width=0.3\textwidth]{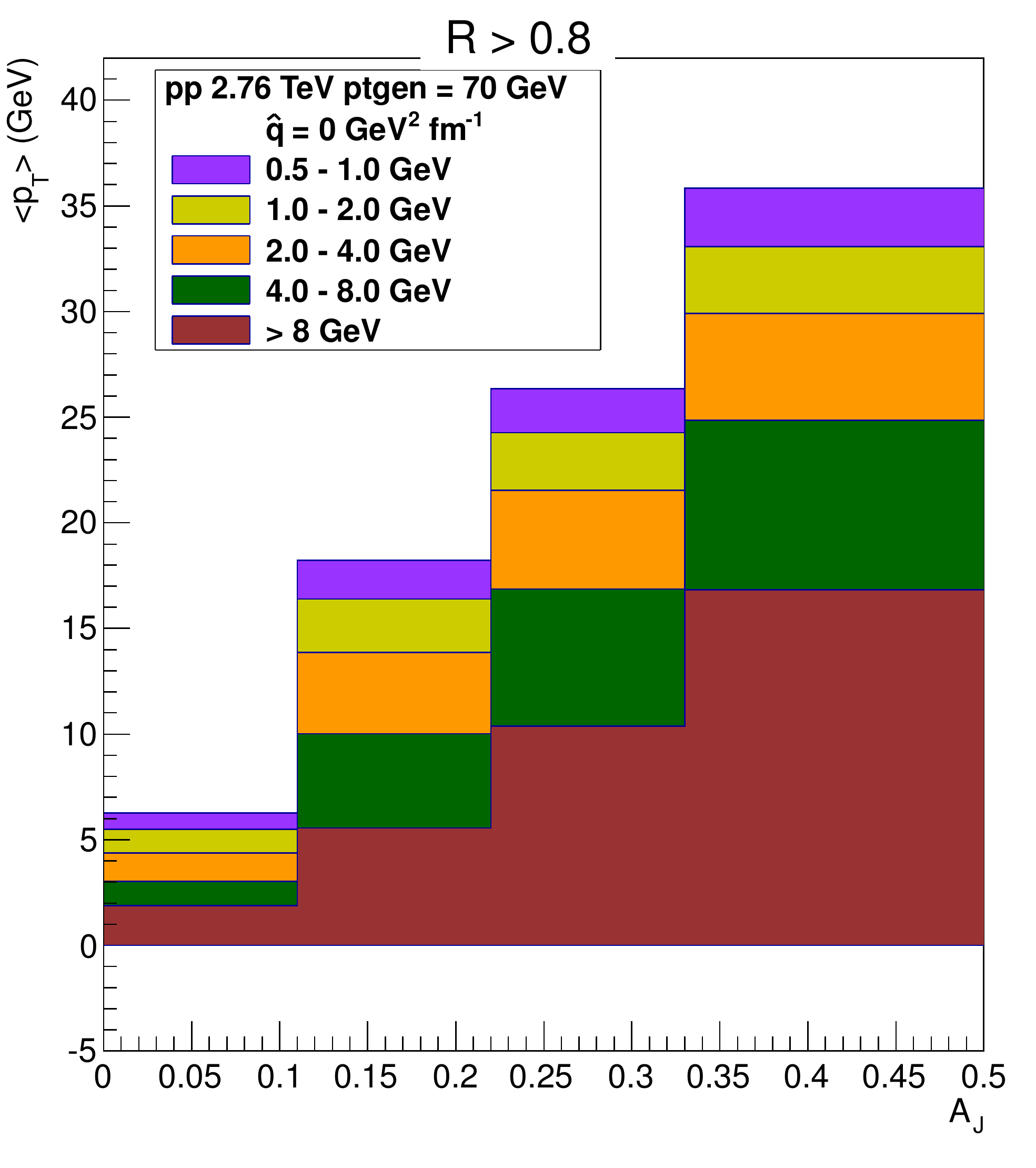}
		\label{fig:MissPt_qhat0_out}
		}
		\subfigure[$< p_T^\parallel >$ with $\hat{q} = 8$ GeV$^2$ fm$^{-1}$.]{
		\includegraphics[width=0.3\textwidth]{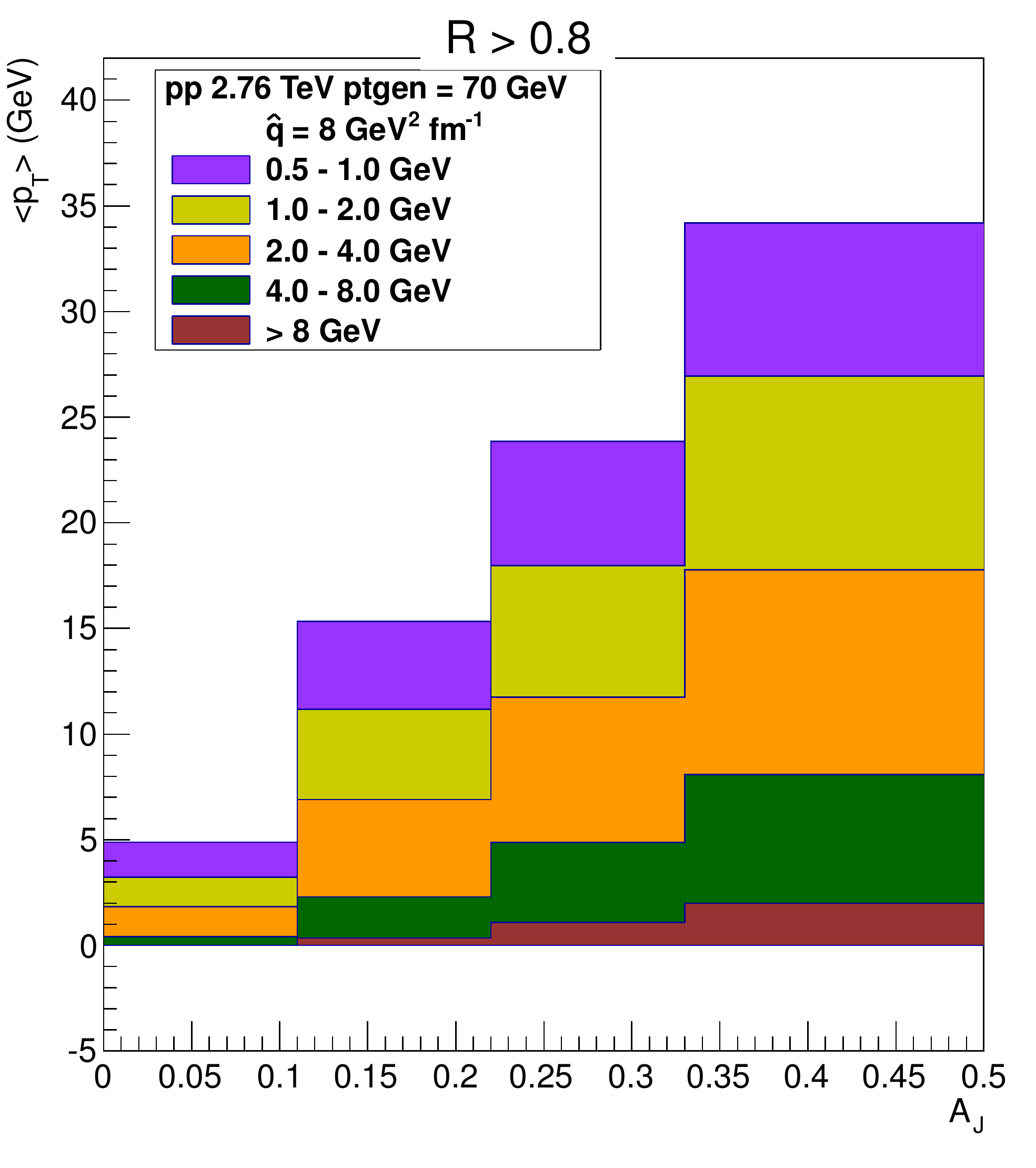}
		\label{fig:MissPt_qhat4_out}
		}
		\vskip-0.3cm
		\caption{Average missing transverse momentum $< p_T^\parallel >$ outside of a cone ($\Delta R > 0.8$) around the leading or subleading jet axis as a function of the dijet asymmetry, $A_J$ for Q-PYTHIA simulation. Colored bands show the contribution to the $< p_T^\parallel >$ for five ranges of particles $p_T$.}
		\label{fig:MissPt}
	\end{center}
\end{figure}

 Now we turn to evaluating the degree of quenching of the data using Q-PYTHIA. For that, we investigate the effect of the influence of the $\hat{q}$ parameter of this model, by changing its value up to 8 GeV$^2$ fm$^{-1}$ while keeping the background fixed to $\sigma = 10.75$ GeV. By increasing $\hat{q}$ in Q-PYTHIA, the jet shower is accelerated with respect to vacuum due to the increasing number of medium-induced splittings. In figure \ref{fig:FJ_Fluct}, one can observe the effect in the dijet asymmetry and azimuthal correlation using jet area method to subtract the background. It seems that a $\hat{q} = 8$ GeV$^2$ fm$^{-1}$ is enough to induce a dijet asymmetry comparable to what is observed in data and the differences between the curve with $\hat{q} = 0$ GeV$^2$ fm$^{-1}$ and $\hat{q} = 8$ GeV$^2$ fm$^{-1}$ for the dijet azimuthal correlation are in agreement with the differences between the CMS Monte Carlo and CMS data.
Moreover, Q-PYTHIA results show the same trend as the average missing transverse momentum data measured by CMS\cite{Chatrchyan:2011sx}, as one can see in figure \ref{fig:MissPt}. Our toy model simulates a system with global variables similar to those measured in the experiments: multiplicity, background level and background fluctuations. However, the distribution of particles in space and the spatial range of their correlations remains unconstrained. We believe that the missing $p_\perp$ observable is extremely sensitive to this last feature, so in the absence of a truly realistic background model, we choose to present only the results for the Q-PYTHIA simulation without background while postponing more detailed studies which require truly realistic backgrounds. As reference we compare the simulations without quenching, like standard PYTHIA ($\hat{q} = 0$ GeV$^2$ fm$^{-1}$) and the CMS simulation for this observable. Our results for the missing transverse momentum outside a cone of $\Delta R = 0.8$ around the leading or subleading jet axis, displayed in figure \ref{fig:MissPt_qhat0_out}, are in well agreement with the CMS simulation. By setting $\hat{q} = 8$ GeV$^2$ fm$^{-1}$ in Q-PYTHIA (figure \ref{fig:MissPt_qhat4_out}) we can compare our simulation with the CMS data. 
Again, we see that, at least qualitatively, it is able to describe the main features of the data, i.e., in general a softest composition of the event, and even outside the cone, the composition is still soft. It seems that a \emph{standard} jet quenching model, where the energy loss is within a close range of the parent parton, is able to explain the soft composition outside the cone, as well as the other two observables, the dijet asymmetry and azimuthal correlation, without a compelling need of a large angle emission mechanism. This can be interpreted as a consequence of event topology since that already in standard PYTHIA, there are events with a large dijet asymmetry ($A_J > 0.3$), which are characterize by a large hard composition outside of the cone (see figure \ref{fig:MissPt_qhat0_out}). This contribution can only come from an event in which the subleading jet emits a hard parton at large angle, originating itself a third semi-hard jet. When propagating through a medium, all jets, including this third jet, will be quenched, and consequently, their initial components are transformed into more soft momentum particles.

\section{Conclusions}
\label{Conclusions}

\par By comparing the two background subtraction techniques we saw that jet energy reconstruction shows sensitivity to fluctuations of the background when using FastJet. In turn, the CMS-\emph{like} method has a dependency on the parameter $E_{Tjets}$ and presents some deviations in the angular reconstruction of the dijet pair. This can be related to the intrinsic structure of the background, indicating that may be needed more than an effective $\rho$ and $\sigma$ to characterize a background. As for the description of the observables with Q-PYTHIA, it seems that this model goes in the same direction than data for the three observables addressed in this study: large jet momentum imbalance and an azimuthal correlation which does not deviate much from the proton-proton case for a reasonable value of $\hat{q}$; and the presence of the higher amount of soft particles at large angle without the need of additional mechanisms of large angle emission.


\noindent \small{\textbf{Acknowledgments:} This work was supported by the European Research Council grant HotLHC ERC-2001-StG-279579; by Ministerio de Ciencia e Innovaci\'on of Spain grants FPA2008-01177, FPA2009-06867-E and Consolider-Ingenio 2010 CPAN CSD2007-00042; by Xunta de Galicia grant PGIDIT10PXIB \-206017PR ; by Funda\c{c}\~{a}o a Ci\^encia e para a Tecnologia of Portugal under projects SFRH/BD/64543/2009 and CERN/FP/116379/2010 (LA); and by FEDER.}





\bibliographystyle{elsarticle-num}
\bibliography{Bibliography.bib}

\end{document}